\documentclass[a4paper,12pt]{article}

\usepackage[a4paper,top=1.75cm,bottom=1.75cm,left=2.5cm,right=2.5cm,marginparwidth=1.75cm]{geometry}

\usepackage{amsmath}
\usepackage{graphicx}
\usepackage{wrapfig}
\usepackage[colorlinks=true, allcolors=blue]{hyperref}
\usepackage{titlesec}
\usepackage{rotating}
\usepackage{chngcntr}
\usepackage{multirow}
\usepackage{amsfonts}
\usepackage{subcaption}
\usepackage[
    backend=biber,
    style=alphabetic,
    sorting=ynt]{biblatex}
\providecommand{\keywords}[1]{\textbf{\textit{Keywords---}} #1}

\title{\textbf{Approximation of supply curves}}
\author{Andr\'es M. Alonso\footnote{Department of Statistics and Institute Flores de Lemus, Universidad Carlos III de Madrid. \mbox{E-mail: andres.alonso@uc3m.es}} \hspace{0.1cm} and Zehang Li\footnote{Department of Statistics, Universidad Carlos III de Madrid. E-mail: zehang.li@alumnos.uc3m.es}}
\date{October, 2023}

\begin{document}
\maketitle

\begin{abstract}
In this note, we illustrate the computation of the approximation of the supply curves using a one-step basis. We derive the expression for the L2 approximation and propose a procedure for the selection of nodes of the approximation. We illustrate the use of this approach with three large sets of bid curves from European electricity markets.
\end{abstract}

%\keywords{Functional approximation; supply curve; electricity market}


\section{Introduction}

In this note, we use the reverse representation of a supply curve which consist on considering the curve as a function in the space (P-rice, Q-uantity) instead of the usual representation (Q-uantity, P-rice). The advantage of the representation (P,Q) over the representation (Q,P) is that all curves are defined in the interval $[0, \infty]$ meanwhile in the (Q,P) representation the curves could have different intervals of definition. Literature in market competitions \cite{MENEZES2012712, Kao2014} has witnessed the effectiveness of the preferred (P, Q) representation, which involves modelling supply curves as functions of prices.

\begin{figure}[h]
\centering
\includegraphics[width=0.75\textwidth]{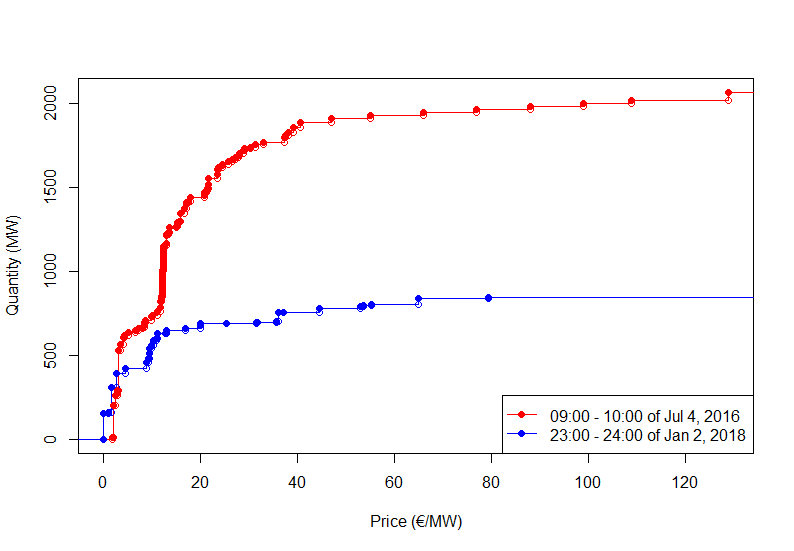}   
\caption{\label{fig:curves}Two curves with a notable difference in the number of steps from Spanish secondary electricity market.}
\end{figure}

A supply curve, $C_t(p)$, is a non-decreasing step function having $n_t$ steps in the positions corresponding to the ordered prices of the offers. Of course, two supply curves could have different number of steps at different positions. For instance, in Figure \ref{fig:curves} we represent two curves from the Spanish secondary electricity market where the curve corresponding to hour 09:00 - 10:00 of July 4, 2016 (in red) has 115 steps while the curve corresponding to hour 23:00 - 24:00 of January 2, 2018 (in blue) has only 32 steps. Given a set of supply curves, $\{C_1, C_2, \ldots, C_T\}$, we could have many positions where some function have a step. A ``natural'' representation is to consider the set of all positions in which some supply function has a step, but the size of that set makes this approach intractable. For instance, in the set of 43848 curves (2016--2020) of the Spanish secondary electricity market we have 16801 different prices, then this ``natural'' representation would be a matrix of dimension 43848 x 16801. These two elements of a set of supply curves, non-decreasing step functions and high dimension, point to the need for a more parsimonious and sufficiently precise representation. 

In \cite{soloviova2021efficient}, a mesh-free interpolation techniques based on radial basis function approximation was proposed for supply and demand curves. However, their approach fails because the resulting approximation is a smooth function (continuous and differentiable) so the step function characteristic is lost. The encoded offer curves (EOC) was proposed by \cite{MESTRE2022122444} which consists on a continuous piecewise version of the true offer curve that approximates the steps. The EOC has the disadvantage of increasing the number of nodes where the curve has changes, which increases the dimension of the representation.

In this note, we will study a proposal based on functions with a single step. We will consider two basis of functions, $\{\phi_{i,n}\}$ and $\{\theta_{i,n}\}$, defined as follows:
\begin{equation} \label{phi}
    \phi_{i,n} = \left\{ \begin{array}{cc}
        0 & \mathrm{if} \ p < p_i \\
        1 & \mathrm{if} \ p \ge p_i
    \end{array}\right.,
\end{equation}
where the set $\{p_1, p_2, ..., p_n\}$ satisfies the conditions $0 = p_1 < p_2 < \ldots < p_n$. 

\begin{equation} \label{theta}
    \theta_{i,n} = \left\{ \begin{array}{cc}
        1 & \mathrm{if} \ p_{i} \le p < p_{i+1} \\
        0 & \mathrm{otherwise}
    \end{array}\right.,
\end{equation}
where the set $\{p_1, p_2, ..., p_n, p_{n+1}\}$ satisfies the conditions $p_1 < p_2 < \ldots < p_n < p_{n+1} = +\infty$. For simplicity, we assume that the set $\{p_i\}_{i=1}^n$ is given, but it is an interesting problem to optimize the grid. 

The rest of the note is organized as follows. Section \ref{sc:L2} presents simple expressions in order to obtain an approximation of a supply curve using the basis of functions, $\{\phi_{i,n}\}$ and $\{\theta_{i,n}\}$. Section \ref{sc:Nodes} deals with the problem of selecting the sets of nodes, $\{p_i\}_{i=1}^n$. Section \ref{sc:Comparison} illustrates the use of this representation with a real dataset. In the Appendix, we provide approximation performance on two additional markets, highlighting the wide applicability across markets and countries.

% {\color{red} For the moment, we eliminated this part but it would be a good idea for you to implement it. "compares its quality with the radial basis representation proposed by \cite{soloviova2021efficient}"}.

\section{Approximation with $\mathcal{L}_2$ loss function} \label{sc:L2}

In order to obtain the approximation, the following loss function will be considered:
\begin{equation}\label{Lr}
    \mathcal{L}_r = || C_t - \widehat{C}_{t,n} ||_r^r = \int_0^{+\infty} \left| C_t(p) - \sum_{i=1}^n c_{t,i,n} \phi_{i,n}(p)\right|^r W(p) \ dp,
\end{equation}
where $r>0$ and $W(p)$ is a non-negative weight function such that $\lim_{p \to +\infty} W(p) = 0$. This condition is needed to guarantee the convergence of the integral (\ref{Lr}).

The approximation of the supply curve is obtained by minimizing the loss function $\mathcal{L}_r$. For a general $r$, this optimization problem will require a computational solution but it is possible to derive an explicit solution for the case $r = 2$. 

The first order conditions of (\ref{Lr}) are as follows:
\[
\begin{array}{cl}
     \frac{\partial \mathcal{L}_2}{\partial c_{t,j,n}} & = -2 \int_0^{+\infty} \left( C_t(p) - \sum_{i=1}^n c_{t,i,n} \phi_{i,n}(p)\right) W(p) \phi_{j,n} \ dp \\
      & = -2 \int_0^{+\infty} C_t(p) W(p) \phi_{j,n} \ dp \quad +2 \sum_{i=1}^n c_{t,i,n} \int_0^{+\infty} \phi_{i,n}(p) \phi_{j,n}(p) W(p) \ dp.
\end{array}
\]

The above expression can be simplified  since $\int_0^{+\infty} C_t(p) W(p) \phi_{j,n} dp = \int_{p_j}^{+\infty} C_t(p) W(p) dp$ and $\int_0^{+\infty} \phi_{i,n}(p) \phi_{j,n}(p) W(p) dp = \int_{\max\{p_i,p_j\}}^{+\infty} W(p) dp$. In what follows, to simplify the notation, we will omit the subscripts $t$ and $n$. Let us denote these integrals by $CW_j$ and $W_j$, respectively. Therefore, we arrive to the following linear system:
\begin{equation} \label{LinearSystem}
    \begin{array}{rcc}
        c_1 W_1 + c_2 W_2 + \cdots + c_{n-1} W_{n-1} + c_n W_n & = & CW_1 \\
        c_1 W_2 + c_2 W_2 + \cdots + c_{n-1} W_{n-1} + c_n W_n & = & CW_2 \\ 
        c_1 W_3 + c_2 W_3 + \cdots + c_{n-1} W_{n-1} + c_n W_n & = & CW_3 \\
        \vdots & & \vdots \\
        c_1 W_{n-1} + c_2 W_{n-1} + \cdots + c_{n-1} W_{n-1} + c_n W_n & = & CW_{n-1} \\
        c_1 W_n + c_2 W_n + \cdots + c_{n-1} W_n + c_n W_n & = & CW_n
    \end{array},
\end{equation}
which can be solved recursively by subtracting one equation from the next. Furthermore, it should be noted that the sequences of coefficients are both decreasing sequences. The solutions of system (\ref{LinearSystem}) are as follows
\[c_1 = \frac{CW_1 - CW_2}{W_1 - W_2},\]
\[c_i = \frac{CW_i - CW_{i+1}}{W_i - W_{i+1}} - \frac{CW_{i-1} - CW_{i}}{W_{i-1} - W_{i}} = \frac{CW_i - CW_{i+1}}{W_i - W_{i+1}} - (c_1 + \cdots + c_{i-1}) \quad \mathrm{for} \ 2 \le i \le n-1\]
and
\[c_n = \frac{CW_n}{W_n} - \frac{CW_{n-1} - CW_{n}}{W_{n-1} - W_{n}} = \frac{CW_n}{W_n} - (c_1 + \cdots + c_{n-1}).\]

Let us consider the summands that appear in the previous solutions, $\frac{CW_i - CW_{i+1}}{W_i - W_{i+1}}$ and $\frac{CW_{i-1} - CW_{i}}{W_{i-1} - W_{i}}$. Using that $C_t$ is a non-decreasing function, we can obtain the following bounds:
\[\frac{CW_i - CW_{i+1}}{W_i - W_{i+1}} = \frac{\int_{p_i}^{p_{i+1}} C(p) W(p) \ dp }{\int_{p_i}^{p_{i+1}} W(p) \ dp} \ge \frac{C(p_i)\int_{p_i}^{p_{i+1}} W(p) \ dp}{\int_{p_i}^{p_{i+1}} W(p) \ dp} = C(p_i),\]
and
\[\frac{CW_{i-1} - CW_{i}}{W_{i-1} - W_{i}} = \frac{\int_{p_{i-1}}^{p_{i}} C(p) W(p) \ dp }{\int_{p_{i-1}}^{p_{i}} W(p) \ dp} \le \frac{C(p_i)\int_{p_{i-1}}^{p_{i}} W(p) \ dp}{\int_{p_{i-1}}^{p_{i}} W(p) \ dp} = C(p_i).\]

Therefore, $c_i \ge 0$ for $2 \le i \le n-1$ and, following the same argument, we can conclude that $c_n \ge 0$. $c_1$ is trivially non-negative since $CW_1 - CW_2$ and $W_1 - W_2$ are non-negatives. Since the coefficients $\{c_i\}$ are non-negative, the obtained approximation, $\widehat{C} = \sum_{i=1}^n c_{i} \phi_{i}$, is also a non-decreasing function which is desirable feature. For a general $\mathcal{L}_r$, we should impose the non-negativeness restrictions for the coefficients, $\{c_i\}$, in the approximation. 

The loss function using the basis $\{\theta_{i,n}\}$ also have a simple expression:
\begin{equation}\label{Lrnew}
\begin{array}{cl}
    \mathcal{L}_r^* & = || C_t - \widehat{C}_{t,n} ||_r^r \ = \int_0^{+\infty} \left| C_t(p) - \sum_{i=1}^n c_{t,i,n}^* \theta_{i,n}(p)\right|^r W(p) \ dp \\
     & = \sum_{i=1}^n \int_{p_{i}}^{p_{i+1}} \left| C_t(p) -  c_{t,i,n}^* \right|^r W(p) \ dp.
\end{array}
\end{equation}
It is clear from the previous expression that $c_{t,i,n}^*$ that minimize $\mathcal{L}_r^*$ does not depend on the other coefficients. 

In the case $r = 2$, the first order conditions are:
\begin{equation}
    \frac{\partial \mathcal{L}_2^*}{\partial c_{t,j,n}^*} = -2 \int_{p_{j}}^{p_{j+1}} \left( C_t(p) - c_{t,j,n}\right) W(p) \ dp = -2 \int_{p_{j}}^{p_{j+1}} C_t(p) W(p) \ dp \quad + \ 2 c_{t,j,n}^* \int_{p_{j}}^{p_{j+1}}  W(p) \ dp,
\end{equation}
which are solved by
\[c_i^* = \frac{\int_{p_{i}}^{p_{i+1}} C(p) W(p) \ dp }{\int_{p_{i}}^{p_{i+1}} W(p) \ dp} = \frac{CW_i - CW_{i+1}}{W_i - W_{i+1}}.\]

It is clear that both basis of functions are closely related, in particular, $\theta_{i,n} = \phi_{i,n} - \phi_{i+1,n}$ for $i = 1, 2, \ldots, n-1$ and $\theta_{n,n} = \phi_{n,n}$ and we can obtain one set of solutions from the other using the following relation $c_i^* = \sum_{j=1}^i c_j$. Since the coefficients $\{c_i\}$ are non-negative, then the coefficients $\{c_i^*\}$ is a non-decreasing sequence. Moreover, the representation $\widehat{C}^* = \sum_{i=1}^n c_{i}^* \theta_{i}$ coincides with the representation $\widehat{C}$.

\section{Selection of the nodes} \label{sc:Nodes}

It is clear that the quality of the approximation depends on the selected set of nodes,  $\{p_i\}_{i=1}^n$. It is expected that a finer grid will lead to a better approximation. However, increasing the number of nodes increases the dimension of the representation. That is, we lose parsimony. In what follows we propose two procedures for selecting the set of nodes. In the next section, these procedures will be compared to a uniform grid on the interval $[0, p_{max}]$ where $p_{max}$ is the maximum observed price.

First, we will consider as reference value of the approximation error the loss function evaluated at $C_t$ and $C_{t-24}$
\begin{equation}\label{Pr}
    \mathcal{P}_{t,r} = || C_t - C_{t-24} ||_r^r = \int_0^{+\infty} \left| C_t(p) - C_{t-24}(p)\right|^r W(p) \ dp.
\end{equation}
Notice that $\mathcal{P}_r$ is a measure of the prediction error of $C_t$ when we use the naive prediction defined by the curve of the previous day at the same time, $t$. 

The goal of the procedures for node's selection is to obtain a mean approximation error 
smaller than the mean prediction error. We divide our sample into a training set,  $\mathbb{C}_1 = \{C_1, C2_2, \ldots, C_{T_1}\}$ and a testing set $\mathbb{C}_2 = \{C_{T_1}, C_{T_1+1}, \ldots, C_{T}\}$. The following steps can be performed in order to choose a set of nodes:

\begin{enumerate}
    \item For each curve in the training set, calculate the prediction error, $\mathcal{P}_{t,r}$. 
    \item Obtain the mean prediction error, $\mathcal{P}_{\bullet,r} = (T_1 - 24) \sum_{t=25}^{T_1} \mathcal{P}_{t,r}$.
    \item Obtain a sample of size $n$ from a reference distribution $F(p)$, $\{p_i\}_{i=1}^n$.
    \begin{enumerate}
        \item As reference distribution, we can use the empirical distribution function of the prices, $\widehat{F}(p) = N^{-1} \sum_{j=1}^N I(p\le p_j)$, where I is the indicator function and $p_j$ are observed prices in the training sample.
        \item Another reference distribution can be derived from the empirical bivariate distribution, $\widehat{F}(p,q) = N^{-1} \sum_{j=1}^N I(p\le p_j, q \le q_j)$, where the pairs $(p_j,q_j)$ are observed bids in the training sample. The reference distribution will be the conditional distribution $\widehat{F}(p | q \ge Q)$, where $Q$ is a selected lower bound. 
    \end{enumerate}
    \item Evaluate the loss function (\ref{Lr}) for the chosen $\{p_i\}_{i=1}^n$ for each curve in the training set.
    \item Obtain the mean approximation error, $\mathcal{L}_{\bullet,r} = (T_1 - 24) \sum_{t=25}^{T_1} \mathcal{L}_{t,r}$.
    \item If $\mathcal{P}_{\bullet,r} > \mathcal{L}_{\bullet,r}$ ends, otherwise increase the number of elements in the set of nodes, $n = n + 1$, and return to step 3.
\end{enumerate}

It is clear that reference distributions, $\widehat{F}(p)$ and $\widehat{F}(p | q \ge Q)$, coincide when $Q = 0$ otherwise they differ. 

In Figure 2 we illustrate the approximations obtained with an increasing number of nodes (5, 10, 15, 20) using the marginal distribution (a) and the conditional distribution (b). It is clear that the approximation improves by increasing the number of nodes. In this example, small differences are obtained between the approximations using the two distributions, $\widehat{F}(p)$ and $\widehat{F}(p | q \ge Q)$.

% \begin{figure}
%      \centering
%      \begin{subfigure}[t]{0.49\textwidth}
%          \centering
%          \includegraphics[width=\textwidth]{figures/Approx_marginal_dist.png}
%          \caption{Approximations using marginal distribution.}
%          \label{fig:Approx_marginal_dist}
%      \end{subfigure}
%      \hfill
%      \begin{subfigure}[t]{0.49\textwidth}
%          \centering
%          \includegraphics[width=\textwidth]{figures/Approx_conditional_dist.png}
%          \caption{Approximations using conditional distribution.}
%          \label{fig:Approx_cond_dist}
%      \end{subfigure}
%         \caption{Approximation comparison by different nodes selection strategies and numbers. The true curve is from  hour 08:00 - 09:00 of August 24, 2018, Spanish secondary market}
%         %\label{fig:Approximations_with_different_nodes}
% \end{figure}

\begin{figure}
     \centering
     \begin{subfigure}[t]{0.49\textwidth}
         \centering
         \includegraphics[width=\textwidth]{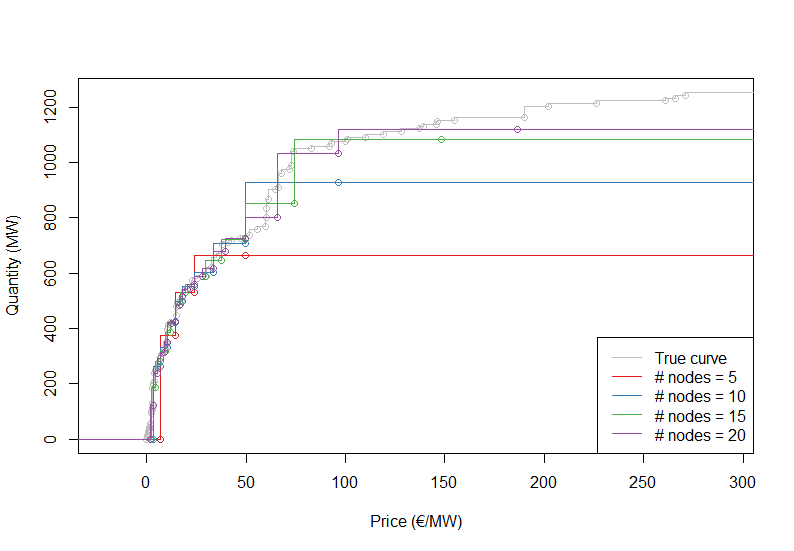}
         \caption{Approximations using marginal distribution.}
         \label{fig:Approx_marginal_dist}
     \end{subfigure}
     \hfill
     \begin{subfigure}[t]{0.49\textwidth}
         \centering
         \includegraphics[width=\textwidth]{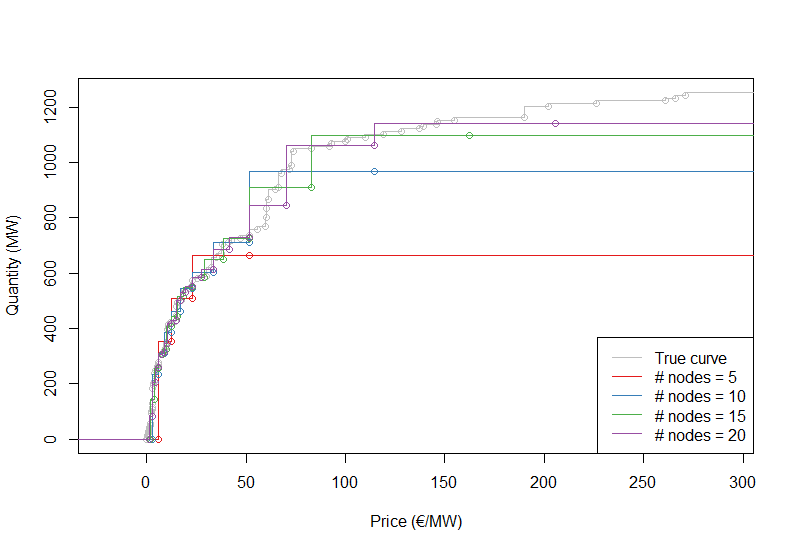}
         \caption{Approximations using conditional distribution.}
         \label{fig:Approx_cond_dist}
     \end{subfigure}
        \caption{Approximation comparison by different nodes selection strategies and numbers. The true curve is from  hour 03:00 - 04:00 of February 24, 2016, Spanish secondary market}
        %\label{fig:Approximations_with_different_nodes}
\end{figure}

\section{Real Data Example} \label{sc:Comparison}

\begin{figure}[h]
\centering
\includegraphics[width=0.75\textwidth]{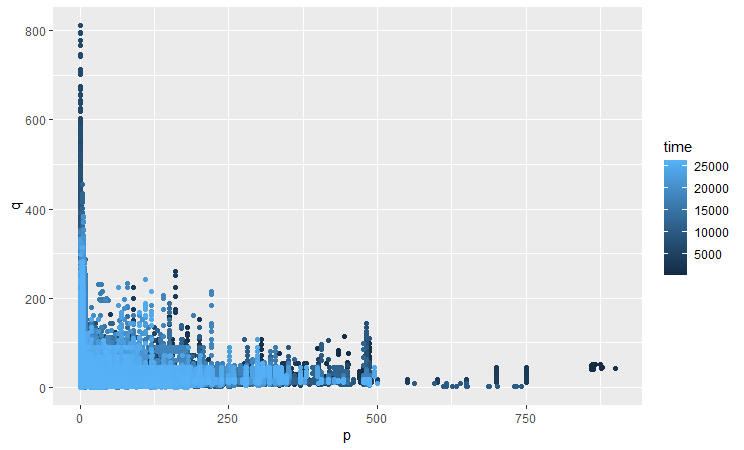}
\caption{\label{fig:bids}Bids pairs (price, quantity) at the Spanish secondary electricity market, 2016-2020.}
\end{figure}

\begin{figure}[h]
\centering
\includegraphics[width=0.75\textwidth]{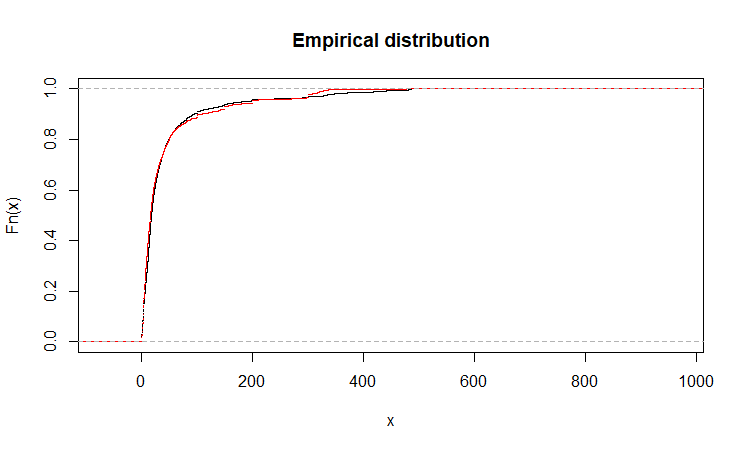}
\caption{\label{fig:distributions}Marginal empirical distribution (black) and Conditional empirical distribution (red).}
\end{figure}

In this section, we compare the performance of the approximation procedure using the data from the Spanish secondary electricity market in the period 2016 -- 2020. We will use the first three years as training sample and the last two as testing samples. In Figure \ref{fig:bids} we represent the bids pairs in the period 2016 -- 2018, and Figure \ref{fig:distributions} represents the reference distribution of steps 3(a) and 3(b), respectively. The black line represents the empirical distribution of the prices and the red line corresponds to the conditional empirical distribution of the prices when the quantities are bigger than 24MWh which is the third quartile of the bid's quantities. It is clear that most of the bids prices are concentrated in the interval from 0 to 100€ for both distributions. The main differences between the distributions appear around 300€, although they are also observed between 100€ and 200€. 

In Table \ref{tab:MApE}, we show the mean approximation error for both reference distributions as well as for an uniform grid that has the same number of points in the set of nodes. The mean scaled approximation error (MScApE) using the naive predictions was $5.9593 \times 10^{-3}$ which is remarkable bigger than the approximation error when we used the proposed procedure. The results show that, from 10 nodes, the procedure that uses the conditional distribution is slightly superior to the procedure that uses the marginal distribution of prices.

 \begin{table}[!h]
 	\centering
 	\begin{tabular}
 		[c]{|l|rr|}\hline
 		Method for Nodes Generation & Nodes number & 1000 * MScApE  \\\hline

		Empirical Distribution	&  5 & 5.1193  \\
		Empirical Distribution	& 10 &  0.8497 \\
		Empirical Distribution	& 15 &  0.3523 \\
		Empirical Distribution	& 20 & 	0.2143 \\ \hline
		Conditional Empirical Distribution	&  5 & 4.5180 \\
		Conditional Empirical Distribution	& 10 &  0.6529 \\
		Conditional Empirical Distribution	& 15 &  0.2954 \\
		Conditional Empirical Distribution	& 20 &  0.1575 \\ \hline
		Uniform Grid &  5 & 157.6908 \\
		Uniform Grid & 10 & 124.4565 \\
		Uniform Grid & 15 & 90.1241 \\
		Uniform Grid & 20 &  65.4651	\\ \hline
	\end{tabular}
	\caption{Comparison of mean scaled approximation errors (MScApE).}%
	\label{tab:MApE}%
\end{table}

% \begin{table}[th]
% 	\centering
% 	\begin{tabular}
% 		[c]{|l|rr|}\hline
% 		Method for Nodes Generation & Nodes number & MApE  \\\hline
% 		Empirical Distribution	&  5 & 40846.18 \\
% 		Empirical Distribution	& 10 &  6779.73 \\
% 		Empirical Distribution	& 15 &  2811.06 \\
% 		Empirical Distribution	& 20 & 	1709.99 \\ \hline
% 		Conditional Empirical Distribution	&  5 & 36048.49 \\
% 		Conditional Empirical Distribution	& 10 &  5209.66 \\
% 		Conditional Empirical Distribution	& 15 &  2357.08 \\
% 		Conditional Empirical Distribution	& 20 &  1257.07 \\ \hline
% 		Uniform Grid &  5 & 1258204.00 \\
% 		Uniform Grid & 10 & 993029.80 \\
% 		Uniform Grid & 15 & 719094.2 \\
% 		Uniform Grid & 20 &  522341.4 \\ \hline
% 	\end{tabular}
% 	\caption{Comparison of mean approximation errors (MApE).}%
% 	\label{tab:MApE}%
% \end{table}

\section{Conclusions}

In this note we propose a simple and parsimonious approximation procedure to a large set of supply curves. The approximation reproduces the two main characteristics of a supply curve: step and non-decreasing functions. We compare our procedure with naive and uniform grid approaches using three dataset from the Spanish and Italian electricity markets and we obtain remarkable smaller mean approximation error. 

An advantage of the proposed method is that it considerably reduces the number of nodes needed to represent a large set of supply curves. Another relevant advantage is that obtaining the approximation is achieved through explicit formulas, which makes the computational cost lower. This lower-dimensional representation and reduced computational cost opens the possibility of exploring its use in classification and supply curve prediction problems as in \cite{ziel2016}.

\printbibliography

\newpage

\appendix

\section{Approximation performance with data from day-ahead markets of Spain and Italy}

The Appendix incorporates the mean approximation error of scaled data (MScApE) obtained from both the Spanish day-ahead market and the Italian market. This is to underscore the robustness of the proposed approach, emphasizing its applicability across diverse markets and countries. To facilitate comparisons, all curves are scaled by the maximum quantity in each training set which is composed of curves from 2016 -- 2018 of different markets. The MScApE using the naive predictions was $1.9838 \times 10^{-3}$ for the Spanish day-ahead market and $3.5101 \times 10^{-3}$ for the Italian day-ahead market. 

\begin{table}[th]
	\centering
	\begin{tabular}
		[c]{|l|rrr|}\hline
		\multirow{5}{*}{\centering Method for Nodes Generation}
        & \multirow{5}{*}{\centering Nodes number}
        & \multicolumn{1}{p{2cm}}{\raggedleft 1000 * MScApE (Spanish day-ahead market)} 
        & \multicolumn{1}{p{2cm}|}{\raggedleft 1000 * MScApE (Italian day-ahead market)} \\\hline

		Empirical Distribution	&  5 & 2.1604  & 3.8683\\
		Empirical Distribution	& 10 &  0.6387 & 0.9555\\
		Empirical Distribution	& 15 &  0.3948 & 0.4029\\
		Empirical Distribution	& 20 & 	0.2763 & 0.2337\\ \hline
		Conditional Empirical Distribution	&  5 & 2.3861 & 3.8054\\
		Conditional Empirical Distribution	& 10 &  0.7301 & 0.9379\\
		Conditional Empirical Distribution	& 15 &  0.3711 & 0.3981\\
		Conditional Empirical Distribution	& 20 &  0.2480 & 0.2299\\ \hline
		Uniform Grid &  5 & 2.0699 & 12.1126\\
		Uniform Grid & 10 & 0.6262 & 3.9601\\
		Uniform Grid & 15 & 0.3990 & 1.8726\\
		Uniform Grid & 20 &  0.2859	& 1.1482\\ \hline
	\end{tabular}
	\caption{Comparison of mean scaled approximation errors (MScApE).}%
	\label{tab:MApEDaily}%
\end{table}

\end{document}